\documentstyle[12pt]{article}

\catcode`\@=11

\def\citerange#1{\if@filesw\immediate\write\@auxout{\string\citation{#1}}\fi
  \def\@citea{}[{\@for\@citeb:=#1\do
    {\@citea\def\@citea{--}\@ifundefined
       {b@\@citeb}{{\bf ?}\@warning
       {Citation `\@citeb' on page \thepage \space undefined}}%
    \hbox{\csname b@\@citeb\endcsname}}}]}

\def\underline#1{\relax\ifmmode\@@underline#1\else
        $\@@underline{\hbox{#1}}$\relax\fi}
\@twosidetrue

\catcode`\@=12

\relax

\newskip\humongous \humongous=0pt plus 1000pt minus 1000pt

\newif\ifdtup

\def\oldreffmt#1{\rlap{[#1]} \hbox to 2\parindent{}}

\def\figfmt#1{\rlap{Figure {#1}} \hbox to 1in{}}

\def\beq{\begin{equation}}
\def\eeq{\end{equation}}
\def\bea{\begin{eqnarray}}
\def\eea{\end{eqnarray}}

\def\bq{\begin{quote}}
\def\eq{\end{quote}}

\def \lta {\mathrel{\vcenter
     {\hbox{$<$}\nointerlineskip\hbox{$\sim$}}}}
\def \gta {\mathrel{\vcenter
     {\hbox{$>$}\nointerlineskip\hbox{$\sim$}}}}

\relax

\def\SM{Standard Model }
\def\SMo{Standard Model}
\def\SMp{Standard Model. }

\def\EW{electro--weak }

\def\phi{\varphi}

\def\addcontentsline#1#2#3{}

\begin{document}

%

{
\def\thefootnote{\fnsymbol{footnote}}
\thispagestyle{empty}

\ \vskip -.8cm
\ \hskip 12.1cm HD--THEP--92--48 rev.

\ \vskip -1.2cm
\ \hskip 12.1cm November 1992

\vskip 2.3cm
\begin{center}
      {\Large\sc\bf Constraints on New Physics from the}\\
                   \     \\
      {\Large\sc\bf Higgs and Top Masses}\\
\vskip 2.0cm
      {U. Ellwanger\footnote{Heisenberg Fellow,
                             Email: I96@VM.URZ.UNI-HEIDELBERG.DE}
     and M. Lindner\footnote{Heisenberg Fellow,
                             Email: Y29@VM.URZ.UNI-HEIDELBERG.DE}}\\

\vskip .8cm
      {\sl  Institut f\"ur Theoretische Physik\\
      der Universit\"at Heidelberg\\
      Philosophenweg 16, D--W--6900 Heidelberg}\\
\end{center}

\vskip 3.0cm
\begin{center}{\Large\bf Abstract}\end{center}
\par \vskip .05in
Triviality and vacuum stability bounds on the Higgs and top quark masses
in a rather general class of supersymmetric extensions of the \SM are
compared with the corresponding bounds without supersymmetry. Due to
generic differences of those bounds we find that experimental knowledge
of the Higgs and top masses may provide a ``pointer'' into one of
these directions. Depending on the values of the masses, however, both
scenarios or none could also be allowed.
}
\newpage
%

\setcounter{page}{1}
\setcounter{footnote}{0}

The \SM of \EW interactions is a quantum field theory which is in agreement
with all existing experimental data. This includes also some evidence for
radiative corrections as required by the theory. Nevertheless it is for
different reasons very likely that the \SM is embedded into a larger
framework. One of the most important reasons is the so called hierarchy
problem which is based on the observation that the quadratic divergences
of the Higgs sector make it hard to explain a big hierarchy between
$v\simeq 175~GeV$ and a very high scale of new physics $\Lambda$. The
hierarchy problem is, however, only a strong argument for new physics
beyond the \SM if the cutoff has a physical meaning. In the renormalizable
\SM itself the problem does not exist since it is absorbed by renormalization.
One might therefore take an extreme attitude and dismiss all those arguments
for new physics.

But even then the ad hoc invention of the Higgs sector in order to break
the \EW symmetry does not necessarily imply that fundamental scalar fields
must exist. Like in the case of the Ginzburg--Landau description of
superconductivity these scalars might turn out to be just an effective
parametrization of some more complex dynamical scenario. However,
independently of the question whether the \SM is just an effective
field theory up to some scale $\Lambda$ the allowed range of parameters
is restricted.  These restrictions stem from the possibility that the
vacuum of the theory can be unstable \cite{unstvac} or that the model is
``trivial'', which means that the only consistent version of the theory
is the free, non--interacting case \cite{triv}. In the language of running
coupling constants these two problems can be phrased as the possibility
that the Higgs self coupling $\lambda(\mu)$ becomes negative such that the
Higgs potential is unbounded from below, or the possibility that one of
the running couplings develops a Landau singularity \cite{Landau}.
Both type of problems can in principle occur at an arbitrarily high
scale $\mu$, but in order to be physically relevant one has to require
that $\mu<\Lambda$, where $\Lambda$ is the range of validity of the
\SMo\footnote{Note, however, that a Landau singularity at the embedding
scale can be considered as an indication of compositness at this scale
\cite{sigm,comp}.}. If the hierarchy problem is not solved in some
unexpected way $\Lambda$ should probably not exceed a few $TeV$. Apart
from the Higgs self coupling $\lambda$ the other possibly large coupling
in the \SM is the top quark Yukawa coupling. Accordingly these restrictions
lead to constraints on the physical Higgs and top quark masses
\citerange{CMPP,Sher}.

Alternatively, if fundamental scalars really exist, a natural solution
to the hierarchy problem is given by supersymmetry. This is because scalars
emerge naturally and quadratic divergences are canceled beyond the
supersymmetry breaking scale $\Delta$, thus the hierarchy problem is solved
if $\Delta\simeq 1~TeV$. The supersymmetric extension of the \SM is,
however, by no means unique. But it is very natural to assume, that any
supersymmetric extension of the \SM is a consistent field theory up to a
GUT or even the Planck scale; after all this possibility is the main
motivation for the introduction of supersymmetry. This implies again the
absence of Landau singularities for the running couplings, now up to these
very large scales. The couplings under consideration are Higgs self
couplings and the top quark Yukawa coupling as before; hence one obtains
again constraints on the physical Higgs and top quark masses. Within a
general supersymmetric extension, however, lower bounds on the lightest
Higgs mass from the condition of vacuum stability cannot be obtained due
to the different form of the scalar potential and the radiative corrections.

Bounds on the  mass of the lightest Higgs scalar in the framework of the
so-called  minimal extension have been discussed in much detail recently
\citerange{Ok,Brig}.  Apart from refs.~\cite{Ell1,ERZ}, however, constraints
from a consistent  ``high energy input'' have not been taken into
consideration in these investigations. Within non--minimal extensions as,
e.g., the addition of a gauge singlet to the Higgs sector
\citerange{Nill,Esp3} these bounds become weaker. Including the leading log
radiative corrections the corresponding upper bounds have recently been
computed in \citerange{Vel,Esp3}.
This latter model can actually be viewed as the  appropriate testing ground
for the general assumption of supersymmetry.  It is sufficiently general and
contains the minimal extension for  special choices of its parameters. The
addition of further doublets to  the Higgs sector would not change the
upper bound on the mass of lightest  Higgs field \cite{Dre,Flor,Esp2}.

A comparison of the constraints on the Higgs and top quark masses within
the \SM and its minimal supersymmetric extension has recently been performed
in \cite{Kras}. There, however, the \SM was assumed to remain valid up to
scales beyond $10^{10} GeV$, and just the minimal supersymmetric extension
was considered. Also the triviality constraint on the top quark Yukawa
coupling was not implemented. In contrast we will use the non--minimal
extension described above, which allows a more general supersymmetric
scenario. Furthermore we believe that in the absence of supersymmetry it
is sensible to require the absence of Landau singularities only up
to a few $TeV$, since the unsolved hierarchy problem will very likely
require such a low embedding scale\footnote{There might, however, exist
solutions of the hierarchy problem {\em within} the \SM which would then
require to take the model serious up to the GUT-- or even the Planck
scale \cite{BW}.}.

Below we will sketch the derivation of the constraints on the Higgs and
top quark masses for the two cases beyond the leading log approximation,
where we make use of results obtained
already elsewhere. From a comparison of these constraints we can learn,
once the Higgs and top masses are experimentally known (or better
constraint), whether perturbative supersymmetry up to $10^{16}$ GeV is
allowed or excluded, or whether an unspecified embedding of the \SM at
a few $TeV$ (or higher) is allowed or excluded. We will discuss
whether some experimental regions can be understood as pointers into
one of those directions.

Within the \SMo, the two undetermined couplings $g_t$ and $\lambda$,
which are related to the unknown top and Higgs masses via $g_t=m_t/v$
and $\lambda=m_H^2/2v^2$, can develop Landau singularities or an
unstable potential even at rather low scales. The renormalization group
flow is given by $dg_t/dt=\beta_t$ and $d\lambda/dt=\beta_\lambda$ where
\beq
16\pi^2\beta_t =
\left( \frac{9}{2}g_t^2 -\frac{17}{12}g_1^2-\frac{9}{4}g_2^2
-8g_3^2\right) g_t~,
\label{betat}
\eeq
and
\beq
16\pi^2\beta_\lambda =
\left( 12\lambda^2 - (A-12g_t^2) \lambda +B -12g_t^4\right)~.
\label{betal}
\eeq
Here $t=\ln{(\mu/\mu_0)}$ and
\beq
A=3g_1^2+3g_2^2~;\quad B=\frac{3}{4}g_1^4+\frac{3}{2}g_1^2g_2^2
+\frac{9}{4}g_2^4~.
\label{AB}
\eeq
{}From the above beta functions we can immediately read off three possible
problems:

\begin{itemize}
\item If $g_t$ is large the running coupling $g_t(\mu)$ can develop a
Landau pole. For large $g_t$ eq.~(\ref{betat}) can be approximated by
$16\pi^2\beta_t = 9/2~ g_t^3$, which leads after integration to the
approximate solution
\beq
\frac{1}{g_t^2(\mu)} = \frac{1}{g_t^2(\mu_0)}
- \frac{9}{16\pi^2} \ln{\left( \frac{\mu}{\mu_0} \right)} ~.
\label{runt}
\eeq
The appearance of a Landau pole in $g_t(\mu)$ (i.e. a zero in
$1/g_t^2(\mu)$) in the physical region below the embedding scale $\Lambda$
is avoided if the top mass is limited by
\beq
\frac{m_t^2}{v^2}=g_t^2(m_t)
<      \frac{16\pi^2}{9\ln{(\Lambda/m_t)}}
\simeq \frac{16\pi^2}{9\ln{(\Lambda/v)}}~,
\label{tlimit}
\eeq
which leads to a bound which is in the \SM weaker than the other two
bounds below. The approximation above describes the true result for
small $\Lambda$ actually quite well. For large $\Lambda$ the bound
(\ref{tlimit}) is too stringent which is immediately clear from the
omission of the gauge couplings in the $\beta$--function. The full
$\Lambda$--dependence of this bound with all running gauge couplings
taken into account was discussed in \cite{Lind}.
\item
For large $m_H$, i.e. large $\lambda$ and small $g_t$ the
$\beta$--function eq.~(\ref{betal}) simplifies and becomes
$16\pi^2\beta_\lambda\simeq 12 \lambda^2$. Integration leads then to
\beq
\frac{1}{\lambda(\mu)} = \frac{1}{\lambda(\mu_0)}
- \frac{3}{4\pi^2} \ln{\left( \frac{\mu}{\mu_0} \right)} ~.
\label{runl}
\eeq
To avoid a Landau pole of $\lambda(\mu)$ in the physical region one must
require
\beq
\frac{m_H^2}{2v^2}=\lambda(m_H)
<      \frac{4\pi^2}{3\ln{(\Lambda/m_H)}}
\simeq \frac{4\pi^2}{3\ln{(\Lambda/v)}}~.
\label{llimit}
\eeq
This approximate ``triviality'' bound for $m_H$ is again quite accurate
for small $\Lambda$ while it is somewhat too stringent for large $\Lambda$.
The full problem has been studied in detail with all effects included
in ref.~\cite{Lind}. Note that this full result has also a weak top mass
dependence.
\item Finally for small $\lambda$ (and moderate $g_t$) the $\beta$--function
eq.~(\ref{betal}) can be simplified to become
$16\pi^2\beta_\lambda\simeq B - 12g_t^4$.
This leads to the approximate solution
\beq
\lambda(\mu) = \lambda(\mu_0) + \frac{B-12g_t^4}{16\pi^2}
\ln{\left( \frac{\mu}{\mu_0} \right) }~.
\label{applam}
\eeq
{}From eq.~(\ref{applam}) one can infer immediately that the solution can
turn negative for $12g_t^4>B$ which would change the sign of the quartic
coupling leading to an unbounded potential. This must be avoided in the
physical region below $\Lambda$.

Eq.~(\ref{applam}) together with
$\lambda(\Lambda) > 0$ and $\lambda(\mu_0) = m_H^2/2v^2$
translates into a lower bound on $m_H$ for large $m_t$. The approximation
leads to
\beq
\frac{m_H^2}{2v^2} > \frac{12m_t^4 - Bv^4}{16\pi^2v^4}
                    ~\ln{\left(\frac{\Lambda}{m_H}\right)}~,
\label{vacstab}
\eeq
which shows how the bound starts at a certain value of $m_t$,  and how
it grows with $\Lambda$. The bound (\ref{vacstab}) is, however, typically
somewhat above the full numerical result \cite{Lind}.  A detailed numerical
study of eq.~(\ref{betal})  with a number of other effects included (such
as newer data, the most important two loop contributions to the
$\beta$--functions, thresholds etc.)  was performed in ref.~\cite{LiSZ}.
Note that $\lambda(\mu)$ in eq.~(\ref{applam})  can become negative
immediately for $\mu$ above $m_H$ if the initial value of $\lambda(m_H)$
goes to zero and if $m_t$ is big enough to change the sign of the
$\beta$--function. This explains the $\Lambda$ independence of this bound
for very small Higgs masses.
\end{itemize}
%
When the three bounds discussed above are combined, we see that the allowed
region in the Higgs--top mass plane is bounded to a $\Lambda$ dependent
range around the origin (see Fig. 2 in \cite{Lind}). Since the development
of Landau pole(s) and of an unstable vacuum can be understood as
``accidents'' of the renormalization group flow it is also intuitively clear
why the bounds are most restrictive for highest $\Lambda$, i.e. the largest
running distance. Even though $\Lambda$ is in principle a free parameter
we know that, for large $\Lambda$, the hierarchy problem will reappear as
soon as we actually specify an embedding of the \SMp Unless some unusual
mechanism solves the hierarchy problem within the \SMo this implies probably
that $\Lambda$ should not be
very large, most likely only a few $TeV$. In that case the bounds become
weak, but they are still very interesting\footnote{Note, however, that it
would still be interesting if these bounds were violated experimentally for
some larger $\Lambda$ since this would establish an experimental upper limit
on the range of validity of the \SMp}. We will include the precise numerical
results for the bounds just discussed in Fig.~\ref{F1} in the comparison
at the end.

As outlined in the introduction, we will also discuss bounds on the Higgs
and top masses within a non--minimal supersymmetric extension of the \SMp
The Higgs sector of this non--minimal extension consists of two Higgs
doublets, $H_1$, $H_2$ and a singlet $S$. It is also motivated by the fact
that it can get along with dimensionless supersymmetric couplings (no
$\mu H_1H_2$ term in the superpotential), so that the \EW scale is
introduced through the soft breaking terms only. (Possible additional
dimensionful couplings will not modify the considerations below.) Since
it is more general than the minimal supersymmetric extension, it is less
restrictive; in particular already the tree level upper bound on the mass
of the lightest Higgs scalar is not given by $M_Z$, but depends -- in some
analogy to the non--supersymmetric model -- on a dimensionless coupling
$\lambda$ \cite{Dre}.

All relevant dimensionless couplings appear in the superpotential in the
form
\beq
W=g_t ~ Q_{L} H_2 T_R + \lambda ~H_1H_2S + \frac{\kappa}{3} S^3~.
\label{formW}
\eeq
Here $Q_{L}$ denotes the doublet containing the left--handed top and bottom
quarks, $T_R$ the right--handed top quark, and the vacuum expectation
value $v_2$ of the Higgs doublet $H_2$ generates a top quark mass
\beq
m_t = g_t v_2~.
\label{mtop}
\eeq
In addition we take the following soft supersymmetry breaking trilinear
couplings and masses into account:
\beq
V_{soft}=(A_t g_t~ Q_{t,L} H_2 T_R+A_\lambda \lambda~ H_1H_2S)+h.c.
+m_1^2|H_1|^2+m_2^2|H_2|^2+m_S^2|S|^2.
\label{Vsoft}
\eeq
Additional terms play no role subsequently. For the derivation of an upper
bound of the lightest Higgs scalar of the model we adopt the following
strategy: We consider the 2 by 2 mass matrix of the scalar neutral $H_1-H_2$
sector and study its lightest eigenvalue, which constitutes such an upper
bound. In this mass matrix we include the leading radiative corrections
induced by top-quark and top-squark loops. Here we neglect the bottom quark
mass and a possible splitting between the top squarks. The contributions of
the gauge and the Higgs sector have been found to affect the final result
only by $\sim 5~GeV$ \cite{Hab,Chan} in the direction of decreasing
the upper bound on $m_H$. (Also in the case of the extended Higgs
sector by the singlet these contributions can be estimated to be numerically
unimportant.) Two loop effects have been found to be of the order of
$\sim 5~GeV$ \cite{Esp1} and
the difference between the pole mass and the second derivative
of the effective potential $\sim 3~GeV$ \cite{Chan,Brig}. Hence we are on
the safe side if we
add $10~GeV$ to our upper bound on $m_H$ obtained below in order
to take these contributions into account. Furthermore we neglect terms of
$O(M_Z^2/m_t^2)$, which point into the lower direction anyway \cite{Hab}.
Now, with the help of the results of \cite{Ell1}, one finds the following
elements of the mass matrix $M^2_{ij}$ after elimination of $m_1^2$ and
$m_2^2$ by means of the extremal equations:
\bea
M_{11} &=& M_Z^2\cos^2(\beta)+\Delta\tan(\beta)-
           \frac{3m_t^4\lambda^2<S>^2}
           {8\pi^2v_2^2m_{sq}^4}\left[A_t+\lambda<S>\cot(\beta)\right]^2~,\\
  \label{m11}
M_{22} &=& M_Z^2\sin^2(\beta)+\Delta\cot(\beta)+                         \\
       & & \frac{3m_t^4}{8\pi^2v_2^2}
           \left[2\ln\left( \frac{m_{sq}^2}{m_t^2}\right)+
           \frac{2A_t(A_t+\lambda<S>\cot(\beta))}{m_{sq}^2}-
           \frac{A_t^2(A_t+\lambda<S>\cot(\beta))^2}{6m_{sq}^4}\right],
                                                                \nonumber  \\
  \label{m22}
M_{12} &=& M_{21}=
           -M_Z^2\sin(\beta)\cos(\beta)-\Delta+\lambda^2w\sin(2\beta)+
                                                    \nonumber \\
       & & \frac{3m_t^4\lambda<S>
           (A_t+\lambda<S>\cot(\beta))}{8\pi^2v_2^2m_{sq}^2}
             \left[ 1-\frac{A_t(A_t+\lambda<S>\cot(\beta))}
             {6m_{sq}^4}\right]~,
  \label{m12}
\eea
with
\beq
\tan(\beta)=\frac{v_2}{v_1}~,
\label{tan}
\eeq
\beq
w=v_1^2+v_2^2\simeq (174~GeV)^2~,
\label{w}
\eeq
\beq
\Delta=\lambda A_\lambda <S>+\lambda \kappa <S>^2-\frac{3m_t^2A_t\lambda <S>}
       {16\pi^2v_2^2}~\ln\left(\frac{m_{sq}^2}{M_Z^2}\right)~.
\label{delta}
\eeq
Actually, neglecting the trilinear couplings $A_t$ and $A_\lambda$, in the
leading log approximation and in the limit $\tan(\beta) \to \infty$ the
following analytic expression for the upper bound on the mass squared
$m_h^2$ of the lightest Higgs scalar can be given \cite{Elw}:
\beq
m_H^2 \leq M_Z^2 \left[ 1 - \sin^2(2\beta) +
      \frac{2\lambda^2}{g_1^2+g_2^2}\sin^2(2\beta)\right] +
      \frac{3}{4\pi^2}v_2^2g_t^4~
      \ln{\left(\frac{m_{sq}^2}{m_t^2}\right)}~.
\label{lighth}
\eeq
We have found numerically, that for non--vanishing $A_t$ and $A_\lambda$
and for arbitrary vacuum expectation value $<S>$ and $\kappa$ the
bound (\ref{lighth}) is exceeded by at most $10~ GeV$ provided $A_t$ and
$A_\lambda$ are bounded by $1~ TeV$ (in agreement with the observation
made in \cite{Ell1}).

{}From eqs.~(\ref{mtop}) and the mass matrix $M_{ij}$ (or the approximate
result (\ref{lighth})) it is evident that upper limits on the couplings
$g_t$ and $\lambda$ turn into upper limits on $m_t$ and $m_H^2$. Upper
limits on $g_t$ and $\lambda$ can be obtained from the assumption that
the running couplings develop no Landau singularities up to a certain scale
$\Lambda$ with, e.g., $\Lambda\simeq 10^{16}~GeV$. These limits have
recently been studied in \citerange{Bin,Esp3}. (According to \cite{Esp2}
and \cite{Vel} the limits of \cite{Bin}, which have been obtained using
analytic approximative solutions for the running couplings, are somewhat
too stringent.) From \cite{Esp2}, e.g., one finds for $g_t$
for general $\kappa$ in eq.~(\ref{formW}),
\beq
g_t  \leq  1.13~,
\label{gtlimit}
\eeq
whereas the upper bound on $\lambda$ varies with $g_t$. For $g_t > .5$
on finds (see also \cite{Ell2})
\beq
\lambda \lta  0.87~.  \label{l2limit}
\eeq
The bound (\ref{gtlimit}) translates into
\beq
m_t\leq 195~GeV~.
\label{tbound}
\eeq
A saturation of this bound implies actually a maximization of $v_2$;
explicitly we have with eq.~(\ref{mtop}) and the bound (\ref{gtlimit})
for fixed $m_t$
\beq
v_2^2 \gta \frac{m_t^2}{(1.13)^2}~,
\label{v2bound}
\eeq
or, with $w = v_1^2+v_2^2$ kept fixed,
\beq
\tan^2\beta \quad \gta \quad \frac{m_t^2}{(1.13)^2 w - m_t^2}~.
\label{tanbound}
\eeq
In the evaluation of the upper bound on $m_H$ according to
eqs.~(\ref{m11}) -- (\ref{m12}) we will make no further assumption
on $\beta$. Then one finds that, for $m_t \gta 130~GeV$, the upper bound
on the lightest eigenvalue of $M_{ij}$ is maximized by minimizing
$\tan^2\beta$.  (Note that the last term on the right hand side  of
eq.~(\ref{lighth}) can be written as  $\frac{3m_t^4 (1+tan^2\beta)}
{4\pi^2w\tan^2\beta} \ln\left(\frac{m_{sq}^2}{m_t^2}\right)$.) Hence,
for $m_t \gta 130~GeV$, we can fix $\beta$ by saturating the  bound
(\ref{tanbound}). This expresses the fact that, for increasing
$m_t$, $v_2$ and hence $\tan\beta$ have to increase in order not to violate
the bound (\ref{gtlimit}). Accordingly, whereas the contributions due to the
radiative corrections increase with $m_t$, $\sin^22\beta$ decreases with
$m_t$. This implies that for large $m_t$ (where  $\tan\beta$ has to be large)
the tree level contribution to $m_H^2$  proportional to $\lambda^2$ becomes
negligible. Thus in  this region the upper bound on the lightest Higgs mass
is the same  as in the minimal model including radiative corrections, which
have  to be computed respecting the bound (\ref{gtlimit}). A corresponding
observation has also been made in \cite{Esp3}, where models with additional
singlets and triplets have been considered (and bounds similar to ours  have
been obtained).

{}From a numerical analysis we find with the upper bounds of \cite{Esp2} for
$\lambda$ and
(\ref{gtlimit}) for $g_t$ that the upper limit on $m_H$
varies between  $140~GeV$ and $165~GeV$\footnote{Here $10~GeV$ have been
added in order to take  care of the neglected effects discussed above.} for
maximal $A_t$, $A_\lambda$ and $m_{sq}$ of $1~TeV$.  The top mass, in turn,
is bounded by $195~GeV$  (\ref{tbound}). The combined limits on $m_H$ and
$m_t$ surround the areas around the origin  in Fig.~\ref{F1} denoted by
{\bf SUSY} and {\bf SM+SUSY}.

In the non--supersymmetric case we show the full numerical solution of the
bound corresponding to (\ref{vacstab}) with
$\Lambda=1~TeV$ due to the unsolved hierarchy problem. The correspondingly
allowed area in Fig.~\ref{F1} is marked by {\bf SM} or {\bf SM+SUSY}.
The two cases differ significantly and lead to areas in  $m_H$--$m_t$
parameter space which are exclusively pointing into one of the two
directions (these areas are labeled {\bf SM} and {\bf SUSY}, respectively).
There are also areas, however, where both or neither of the scenarios are
acceptable. These areas are denoted in Fig.~\ref{F1} by {\bf SM+SUSY} and
{\bf NEITHER}.

As already mentioned it is in principle conceivable that there exists a
solution to the  hierarchy problem without involving supersymmetry. In
that case one could make $\Lambda$ within the \SM scenario very large, for
example $10^{15}~GeV$. This would imply that  the bounds for the \SM
scenario would become significantly stronger. We have included in
Fig.~\ref{F1} these stronger \SM bounds as a weak solid line labelled
{\bf $10^{15}$}. Note that the area labeled {\bf NEITHER} would then grow
significantly and the area labeled {\bf SUSY} exclusively would also grow,
while the pure {\bf SM} range as well as the {\bf SM+SUSY} range would
shrink. We have also included the experimental lower bounds on the Higgs
mass \cite{minhiggs} and the top mass \cite{mintop} as dashed--dotted line.

The bounds of ref.~\cite{Kras} would be obtained if we would restrict our
discussion to the minimal supersymmetric scenario, if we would ignore
simultaneously the bound on $m_t$ from eq.~(\ref{tbound}), and if we took
$\Lambda=10^{15}~GeV$ for the \SM bounds without consideration of the
hierarchy problem. We think, however, that it is better to take our more
general scenario as the testing ground of supersymmetry, to implement the
triviality constraint on $g_t$ as well, and that one
should most likely take $\Lambda\simeq 1~TeV$ for the non--supersymmetric
scenario. Consequently our bounds differ significantly from those of
ref.~\cite{Kras}.

\bigskip

{\bf Acknowledgement}

We like to thank W. ter Veldhuis and M. Quir\'os for correspondence on
the couplings within the nonminimal supersymmetric model.
%
%

%

\vskip .7cm

\def\listfigurename{Figure Captions}
\listoffigures

\begin{figure}[htb]
   \vspace{0.1cm}
   \caption{
      Combination of Higgs and top mass bounds of our general
      supersymmetric scenario
      (the fat solid line connecting $(m_H,m_t)=(113~GeV,0~GeV)$
      with $(m_H,m_t)=(0~GeV,174~GeV)$)
      with the \SM bounds for $\Lambda=1~TeV$
      (the fat solid line starting at $(m_H,m_t)=(0~GeV,85.6~GeV)$).
      The resulting four areas are labeled with {\bf NEITHER},
      {\bf SM}, {\bf SUSY} and {\bf SM+SUSY} respectively
      to indicate the allowed scenario(s).
      The weak solid line shows the stronger \SM bound for
      $\Lambda=10^{15}~GeV$.
      Experimental lower limits for the Higgs and top masses
      are shown as weak dashed--dotted lines.}
   \label{F1}
\end{figure}

\end{document}